\documentclass[aps,prl,preprint,groupedaddress]{revtex4-2}

\usepackage{graphicx}
\usepackage{siunitx}
\usepackage{chemformula}
\usepackage{amsmath}
\usepackage[english]{babel}
\usepackage[nottoc]{tocbibind}
\usepackage[printonlyused,withpage]{acronym}

\begin{document}

\preprint{}

\title{Temperature dependent study of the spin dynamics of coupled \ch{Y3Fe5O12}/\ch{Gd3Fe5O12}/\ch{Pt} trilayers}


\author{Felix Fuhrmann}
\email[]{fefuhrma@students.uni-mainz.de}
\author{Sven Becker}
\author{Akashdeep Akashdeep}
\affiliation{Institute of Physics, University of Mainz, Staudingerweg 7, Mainz 55128, Germany}
\author{Zengyao Ren}
\affiliation{Graduate School of Excellence “Materials Science in
	Mainz” (MAINZ), Staudingerweg 9, Mainz 55128, Germany}
\affiliation{Institute of Physics, University of Mainz, Staudingerweg 7, Mainz 55128, Germany}
\author{Mathias Weiler}
\affiliation{Fachbereich Physik and Landesforschungszentrum OPTIMAS, Rheinland-Pfälzische Technische Universität Kaiserslautern-Landau, 67663 Kaiserslautern, Germany}
\author{Gerhard Jakob}
\affiliation{Institute of Physics, University of Mainz, Staudingerweg 7, Mainz 55128, Germany}
\author{Mathias Kläui}
\affiliation{Institute of Physics, University of Mainz, Staudingerweg 7, Mainz 55128, Germany}
\affiliation{Graduate School of Excellence “Materials Science in
	Mainz” (MAINZ), Staudingerweg 9, Mainz 55128, Germany}
\affiliation{Center for Quantum Spintronics, Norwegian University of Science and Technology, Trondheim 7491, Norway}

\date{\today}

\begin{abstract}
In this study, we investigate the dynamic response of a \ch{Y3Fe5O12} (YIG)/ \ch{Gd3Fe5O12} (GdIG)/ Pt trilayer system by measurements of the ferromagnetic resonance (FMR) and the resulting pumped spin current detected by the inverse spin Hall effect. This trilayer system offers the unique opportunity to investigate the spin dynamics of the ferrimagnetic GdIG, close to its compensation temperature. We show that our trilayer acts as a highly tunable spin current source. Our experimental results are corroborated by micro-magnetic simulations. As the detected spin current in the top Pt layer is distinctly dominated by the GdIG layer, this gives the unique opportunity to investigate the excitation and dynamic properties of GdIG while comparing it to the broadband FMR absorption spectrum of the heterostructure.
\end{abstract}


\maketitle

\section{Introduction}
In recent years, the field of magnonics has been growing. Encoding the information by spin angular momentum instead of moving charge carriers can potentially decrease energy consumption \cite{brataas_spin_2020,chumak_magnon_2015}. This fuels interest in developing magnonic devices, which can be used for magnon logic operations and offer potentially increased speed and lower power consumption \cite{brataas_spin_2020,cramer_magnon_2018}.
Rare earth garnets like \ch{Y3Fe5O12} (YIG) offer a unique platform with long-distance magnon propagation, enabled by its low Gilbert damping constant of down to $\alpha \approx 10^{-5}$ \cite{wei_giant_2022,serga_yig_2010,schmidt_ultra_2020, cornelissen_long-distance_2015}. \ch{Gd3Fe5O12} (GdIG) is a compensated ferrimagnetic rare earth garnet with a temperature-dependent net magnetization that vanishes at the magnetic moment compensation temperature $T_{\mathrm{Comp}} \approx \SI{295}{\kelvin}$ \cite{pauthenet_spontaneous_1958}. Heterostructures of these materials provide an interesting static magnetic system and allow to study the spin dynamics  of the coupled heterostructure \cite{becker_magnetic_2021, gomez-perez_synthetic_2018, roos_magnetization_2022}. 
Currently, the study of antiferromagnets is also an active research area, as it promises materials with resilience against external magnetic fields, long-distance spin transport and naturally high resonance frequencies \cite{lebrun_tunable_2018, baierl_terahertz-driven_2016}. Antiferromagnet-ferromagnet heterostructures \cite{al-hamdo_coupling_2023} and ferrimagnetic systems, especially close to their compensation temperature, provide an interesting platform to study antiferromagnetic (and antiferromagnetically coupled) spin dynamics with more accessible magnetic properties \cite{liensberger_exchange-enhanced_2019, klingler_spin-torque_2018, stanciu_ultrafast_2006}. Until now, the investigation of the dynamics of ferrimagnets close to their compensation temperature has been challenging \cite{ng_survey_2022,li_unconventional_2022}. We show that the coupling to a second layer can be used to facilitate such studies. 

In this study, we experimentally investigate a YIG/GdIG/Pt thin-film heterostructure schematically depicted in Fig. \ref{fig:Fig1} a). We observe a strong impact of the magnetic configuration of our heterostructure on spin dynamics, spin pumping and spin Seebeck effect. We show that the generated spin current originates in the GdIG layer, which gives us the unique opportunity to investigate the GdIG spin dynamics, aided by the coupling to the YIG layer, close to the compensation temperature. This temperature range is usually more difficult to study because of the diverging linewidth of single GdIG layers at $T_{\mathrm{Comp}}$ \cite{ng_survey_2022}. Furthermore, we study the spin current generation in our system, which is tunable by temperature, external field and relative layer thickness \cite{becker_magnetic_2021}.
We drive the ferromagnetic resonance (FMR) modes of our heterostructure and measure the spin current which is pumped across the GdIG/Pt interface when the resonance condition is satisfied \cite{kittel_introduction_2005, tserkovnyak_enhanced_2002}. This spin current, resulting from the spin pumping (SP), is detected by means of the inverse spin Hall effect (iSHE) in the Pt top layer \cite{ sinova_spin_2015} in the experimental configuration sketched in Fig \ref{fig:Fig1} b). We gain more information about the switching behavior of our GdIG layer by observing the spin Seebeck effect (SSE) \cite{geprags_origin_2016}. The SSE measurements are performed by applying an out-of-plane thermal gradient as depicted in Fig. \ref{fig:Fig1} c). This geometry is referred to in literature as the longitudinal spin Seebeck effect (LSSE) \cite{uchida_observation_2010}. We compare these results with SSE measurements, in which the gradient is generated by microwave heating during the SP measurements \cite{schreier_spin_2016}.
The microwave-induced SSE is helpful as a measure to determine the switching of the top GdIG layer during the SP measurement itself, and is less susceptible to temperature mismatch compared to remounting the sample in another setup.

\section{Sample characteristics}
The investigated sample is a trilayer of YIG, GdIG, and Pt, grown on a GGG (001) substrate. The thicknesses of YIG and GdIG are chosen to be \SI{36}{\nano\meter} and  \SI{30}{\nano\meter}  respectively. The growth of the sample via pulsed laser deposition was optimized and the coupling between YIG and GdIG moments was investigated in Ref. \cite{becker_magnetic_2021}. 

\begin{figure}[htb]
	\centering
	\includegraphics[width=0.6\textwidth]{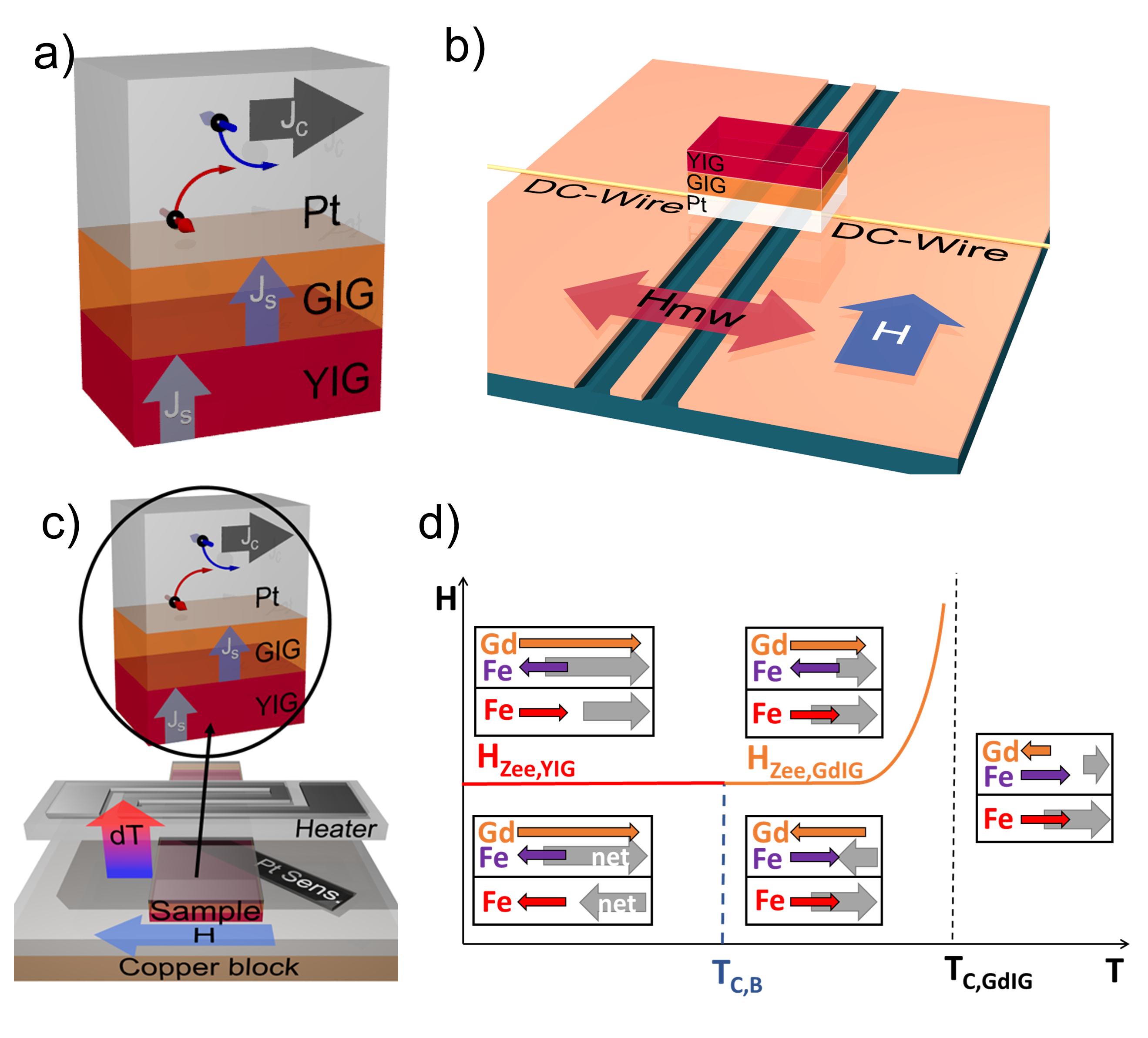}
	\caption{a) Illustration of the YIG / GdIG / Pt trilayer without the GGG substrate. The spin current $J_\mathrm{S}$ which is generated potentially by both, the YIG and GdIG layers, is converted into a charge current $J_\mathrm{C}$ in the Pt layer via the iSHE. b) Illustration of the sample placement on a CPW. The sample is connected with two thin wires, left and right. The voltage $V_{\mathrm{iSHE}}$, generated in the Pt layer of the sample is measured by a Lock-In technique. c) Illustration of the SSE sample stack with a constant temperature gradient. The temperature gradient is established between the Pt-heater on top of the sample and the copper piece (in contact with the VTI of the cryostat). The gradient is estimated by comparing the resistance of the Pt-heater and Pt-sensor below the sample. The zoomed section illustrates the iSHE in the Pt after a spin current is channeled into the Pt top layer. d) Illustration of the YIG / GdIG / Pt trilayer alignment and switching field dependent on temperature. Orange arrows depict the Gd sublattice, and the red (purple) arrows illustrate the direction of the combined Fe-sublattices of YIG (GdIG). The $H_{\mathrm{Zee}}$ line indicates the rough temperature dependence of the magnetic field necessary to switch the respective layer.} 
	
	\label{fig:Fig1} 
\end{figure}

The relative alignment of the YIG and GdIG layer magnetizations was investigated previously by SQUID magnetometry and spin Hall magnetoresistance \cite{becker_magnetic_2021, gomez-perez_synthetic_2018}. With this information and the (microwave-induced) SSE measurements, we can determine the relative alignment in our sample. Figure \ref{fig:Fig1} d) shows an illustration of the switching field (and thus relative alignment) versus temperature for the YIG \SI{36}{\nano\meter}/GdIG \SI{30}{\nano\meter}/Pt \SI{4}{\nano\meter} sample. For temperatures below the bilayer compensation temperature $T_{\mathrm{C,B}}$, the net moment of the GdIG layer $M_{\mathrm{net,GdIG}}$ is larger than the one of the YIG layer $M_{\mathrm{net,YIG}}$. At $T_{\mathrm{C,B}}$, the two layers have the same net moment, which are antiferromagnetically coupled via the Fe-Fe sublattices \cite{becker_magnetic_2021, gomez-perez_synthetic_2018} and thus compensate each other. For temperatures above $T_{\mathrm{C,B}}$ and below $T_{\mathrm{C,GdIG}}$ (the compensation temperature of single layer GdIG), the net moment of the YIG layer $M_{\mathrm{net,YIG}}$ is larger than that of the GdIG layer $M_{\mathrm{net,GdIG}}$. Under small external applied in-plane magnetic fields, the net magnetization is parallel to the applied field while the individual layer magnetizations are antiparallel. For lager external magnetic fields, the Zeeman energy exceeds the exchange coupling, and the layer with the smaller net moment also switches so that both magnetizations are parallel with the external magnetic field. While the net moments $M_{\mathrm{net,GdIG}}$ and $M_{\mathrm{net,YIG}}$ are antiferromagnetically coupled below $T_{\mathrm{C,GdIG}}$, the two layer magnetizations $M_{\mathrm{net,GdIG}}$ and $M_{\mathrm{net,YIG}}$ are always parallel above $T_{\mathrm{C,GdIG}}$.
The results of this trilayer structure are compared to the FMR data of a single GdIG layer of a comparable thickness of \SI{30}{\nano\meter} and a YIG \SI{36}{\nano\meter}/Pt \SI{4}{\nano\meter} bilayer film.

\section{Results and Discussion}
\subsection{Ferromagnetic resonance absorption}
\begin{figure}[htb]
	\centering
	\includegraphics[width=\textwidth]{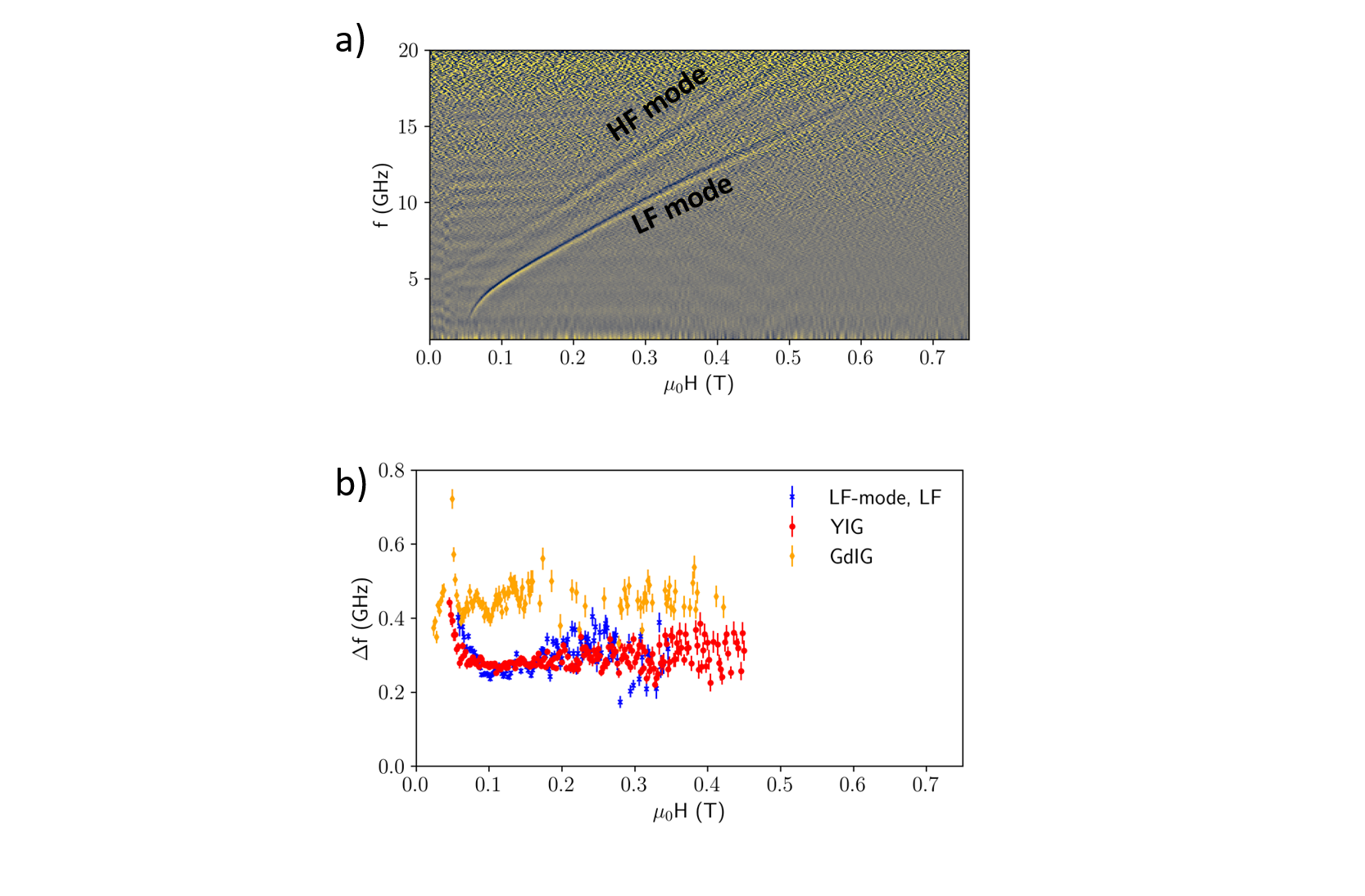}
	\caption{ a) Broadband VNA-FMR measurement at \SI{50}{\kelvin}. The raw data is processed by derivative divide and an fft-filter. The resonance linewidth is extracted by fitting to the derivative of the susceptibility for N resonances \cite{schneider_spin_2007}. b) The plot shows the linewidth at \SI{50}{\kelvin} from the  YIG \SI{36}{\nano\meter}/ Pt \SI{4}{\nano\meter} and GdIG \SI{30}{\nano\meter} samples (red and orange respectively) and the trilayer (blue) in comparison.}
	\label{fig:Fig2}
\end{figure}

To obtain the colormaps in Fig. \ref{fig:Fig2} a) we use a Vector Network Analyzer (VNA) to measure the microwave absorption of the sample. In this case, the frequency is swept by the VNA for each external magnetic field step $\mu_0 \boldsymbol{H}$, resulting in a broadband FMR measurement (bbFMR) \cite{maksymov_broadband_2015}. The obtained raw data is then processed by the derivative Divide (dD) algorithm following Maier-Flaig \textit{et al.} \cite{maier-flaig_note_2018}. 
\begin{align}
	d_\mathrm{D}S_{21} = & \frac{S_{21}(\omega, H_0+\Delta H_\pm)-S_{21}(\omega, H_0 - \Delta H_\pm)}{S_{21}(\omega, H_0) \Delta H_\pm} \\
	\approx & -i \omega A' \frac{d\chi}{d\omega}  
	\label{dD_Formula}
\end{align}
Here, $d_\mathrm{D}S_{21}$ is thus effectively the normalized derivative of the $S_{21}$-parameter (transmission parameter) with respect to magnetic field \cite{maier-flaig_note_2018}. 


In our FMR measurements, we compare VNA-FMR colormaps recorded at different temperatures. We measured our single-layer YIG and GdIG films first, to compare them later to the features of our heterostructure (Fig. \ref{fig:Fig2} a), b)).

For the heterostructure, the signal differs strongly in its shape and temperature dependence from those of the single layers. For a temperature of \SI{50}{\kelvin} we observe two distinct modes, one at lower and one at higher frequencies (Figure \ref{fig:Fig2} a). Both modes do not behave as one would expect from the Kittel equation for in-plane applied external magnetic fields for single layers. However, the linewidth and signal strength of the mode at higher frequencies (HF mode) is more compatible with the GdIG single layer (Fig. \ref{fig:Fig2} b)). The increase of the slope of the mode towards lower temperatures was also previously observed for GdIG \cite{funada_low_2022}. The lower mode (LF mode) resembles the behavior of the single YIG layer.
However, for temperatures in which the two modes are close to each other, it becomes clear that the behavior is increasingly complex for the bilayer system.
To identify the features and investigate their origin in more detail, we complement the FMR data with the SP and SSE measurements.

\begin{figure}[htb]
	\centering
	\includegraphics[width=\textwidth]{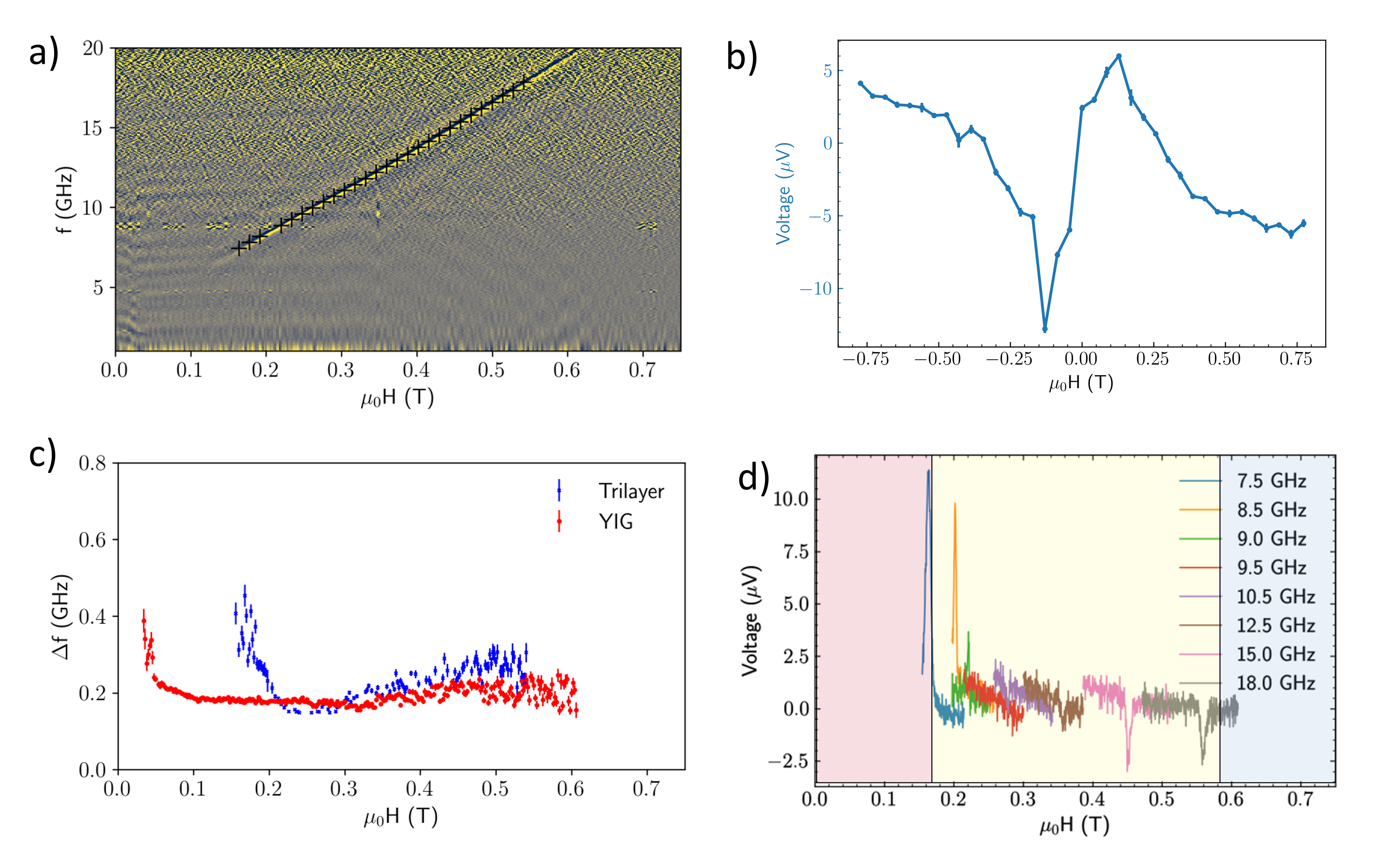}
	\caption{ a) Broadband VNA-FMR measurement at \SI{200}{\kelvin}. The raw data is processed by derivative divide and an fft-filter, which filters signals by frequency apart from an allowed window. The resonance linewidth is extracted by fitting to the derivative of susceptibility for N resonances \cite{schneider_spin_2007,al-hamdo_coupling_2023}. b) The plot shows the SSE measurement by microwave heating at \SI{200}{\kelvin} of the trilayer. c) Extracted linewidth at \SI{200}{\kelvin} from the trilayer and the YIG sample in comparison. d) Measurement of $V_{\mathrm{iSHE}}$ from SP at resonance with different excitation frequency at \SI{200}{\kelvin}. The background colors of the plot red (yellow and blue) refer to the ranges in which the GdIG layer is antiparallel (switching and parallel) to the magnetic field.}
	\label{fig:Fig3}
\end{figure}

\subsection{Spin pumping at ferromagnetic resonance}
Further inspection of the FMR spectrum and generation of spin currents was performed by the observation of the inverse spin Hall effect (iSHE) in the Pt top layer \cite{tserkovnyak_enhanced_2002,sinova_spin_2015}. At the resonance condition, a spin current is pumped from the YIG/GdIG bilayer into the adjacent Pt layer. This generated spin current is dominated by the spin dynamics in the GdIG layer, which offers a chance to investigate the GdIG layer in further detail. While the FMR measurements are sensitive to both the GdIG and YIG spin dynamics, the SP and SSE measurements are complementary, as they only reflect the GdIG spin dynamics.
The investigation of magnetic field sweeps, especially at the switching fields of the respective layers offers the opportunity to study the spin current origin.
The sample is placed on top of the CPW with a thin insulating tape, such that there is no electrical contact from the CPW to the Pt film. The contacts are made by a thin copper wire and silver paste to fix the wire on the Pt layer. 

\begin{figure}[htb]
	\centering
	\includegraphics[width=\textwidth]{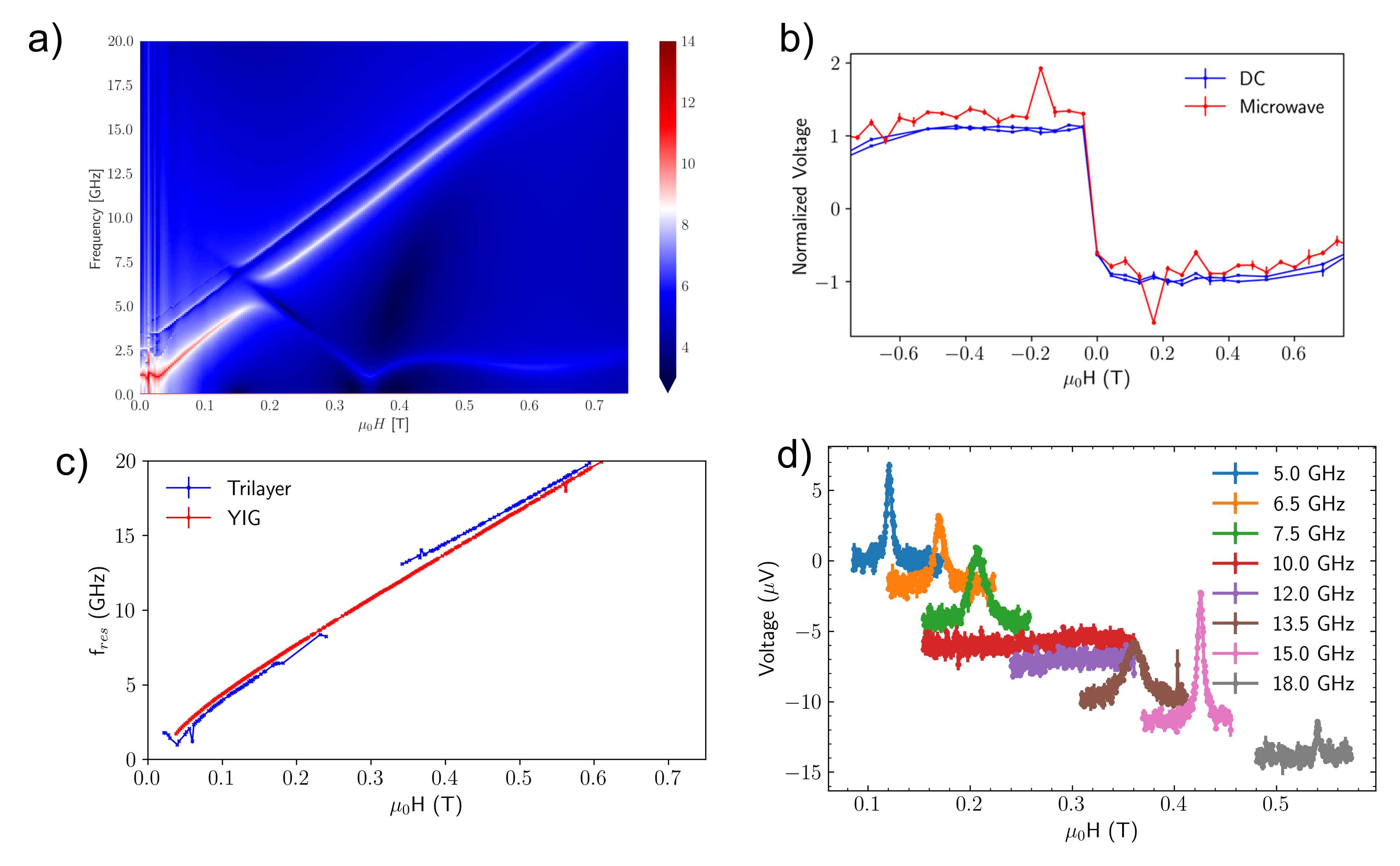}
	\caption{a) FMR spectrum of the micromagnetic simulation. The magnetization is estimated from literature to match a temperature of \SI{250}{\kelvin}. The raw data is processed by derivative divide. b) DC-Spin Seebeck effect measurement, indicating the switching of the top GdIG layer above \SI{0.4}{\tesla}. c) Extracted resonance frequency $f_{\mathrm{res}}$ at \SI{250}{\kelvin} for the trilayer and a single YIG layer in comparison.
	d) Measurement of $V_{\mathrm{iSHE}}$ from SP at resonance with different excitation frequency at \SI{200}{\kelvin}. A constant offset of \SI{2}{\micro\volt} is used to separate the signal at different frequencies visibly in the plot.}
	\label{fig:Fig4}
\end{figure}

The alternating current is generated by an MW-sweeper, which is amplified and fed into the CPW, generating an alternating magnetic field at the sample. The alternating magnetic field is modulated in its amplitude by a frequency of \SI{30}{\hertz} (DC). The resulting voltage $V_{\mathrm{iSHE}}$ is measured in the Pt layer by a Lock-In technique with a Stanford Research (SR) 830 with a SR-530 pre-amplifier, improving the signal-to-noise ratio and removing artifacts in the signal due to small temperature fluctuations inside the cryostat.
The measurements are performed as field sweeps with a constant frequency of the alternating field, at different temperatures. At a temperature of \SI{200}{\kelvin} we can observe a sign-change of the generated signal for a sweep from low to high magnetic fields (lower and higher resonance frequencies) (Fig. \ref{fig:Fig3} d). In this temperature range ($T_{\mathrm{C,B}} < T < T_{\mathrm{C,GdIG}}$), the net moment of the YIG layer is larger than the GdIG net magnetic moment (see Fig. \ref{fig:Fig1} d)). Thus, at low fields, the YIG layer magnetization is parallel to the magnetic field, while the GdIG magnetization is oriented antiparallel. By comparing the sign change to the indicated ranges in Figure \ref{fig:Fig3} d), one can observe the voltage sign at resonance following the GdIG orientation. 
It is noteworthy, that the FMR-spectrum shows a clear resonance line during the GdIG switching (compare Fig. \ref{fig:Fig3} a) and b)). This suggests, that the YIG can still be excited during the GdIG switching, and the FMR spectrum is dominated by the YIG layer in this field range. This is supported by the narrowing of the linewidth during the GdIG switching ($\approx$ \SI{0.2}{\tesla} to \SI{0.5}{\tesla}) (see \ref{fig:Fig3} c)) From the SSE measurements it is clear that the switching is not abrupt, but occurs over an extended field range. In this range, the linewidth is compatible with the linewidth measured in the single YIG layer.
It appears that we are not sensitive to a spin current originating from YIG in this range, as there is no signal of the expected linewidth and sign observed in the $V_ {\mathrm{iSHE}}$ signal. From SSE measurements we can estimate the 50 \% switching field value. For GdIG switched by 50 \% (at $\approx \SI{ 0.3}{\tesla}$) we do not see any distinct peak, supporting the assumption that there is no significant transmitted spin current from the YIG.
The sign change of the voltage signal can be explained easily by a switching of the GdIG layer magnetization. As the magnetization direction of the spin current source (the GdIG layer) inverts, the spin current polarization changes, leading to an inverted voltage $V_{\mathrm{iSHE}}$ (Fig. \ref{fig:Fig3} d)). 

\begin{figure}[htb]
	\centering
	\includegraphics[width=\textwidth]{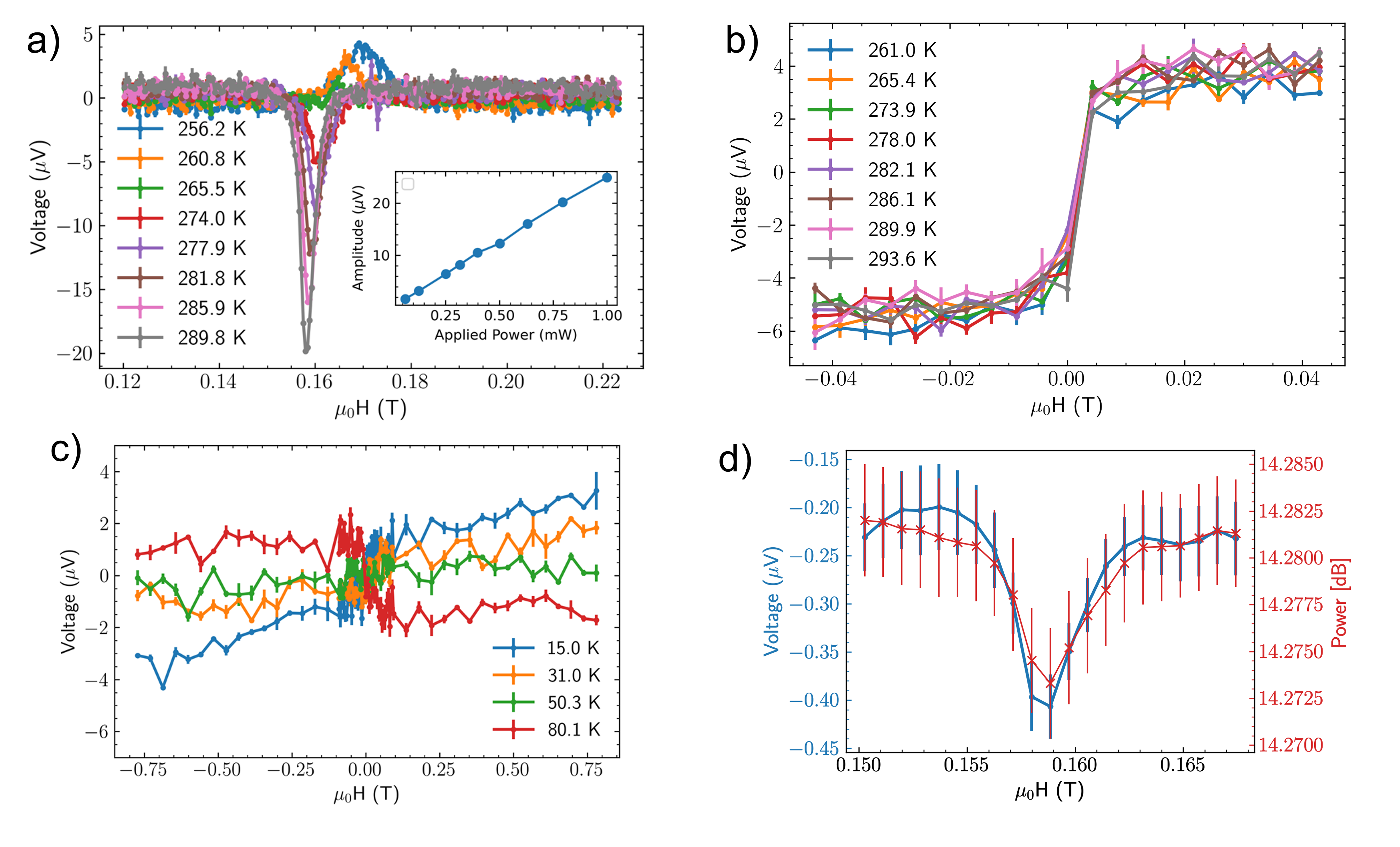}
	\caption{a)  $V_{\mathrm{iSHE}}$ at fixed resonance frequency of \SI{6.5}{\giga\hertz} for different temperatures. The temperature range reaches across the GdIG compensation temperature. The sign of $V_{ISHE}$ switches at $T \approx$ \SI{ 270}{\kelvin}. b) The low-field step of $V_{\mathrm{iSHE}}$ has it's origin in the spin Seebeck effect. The observed SSE signal shows no sign change in contrast to the SP signal over the same temperature range as in a). c) $V_{\mathrm{iSHE}}$ of the SSE by mircowave heating for low temperatures \cite{geprags_origin_2016,ohnuma_spin_2013,cramer_magnon_2017}. d) Microwave absorption (red curve) and $V_{\mathrm{iSHE}}$ (blue curve) in comparison at \SI{295}{\kelvin} and at \SI{6.5}{\giga\hertz}. }
	\label{fig:Fig5}
\end{figure}

For a temperature sweep across $T_{\mathrm{C,GdIG}}$ one can also expect a sign change. The GdIG net moment is inverted across the compensation temperature, as the combined moment of the Fe sublattices exceeds the Gd sublattice moment. Thus, the net magnetization direction of GdIG is inverted, leading to an inversion of the iSHE voltage (see Fig. \ref{fig:Fig5} a)). This again signals, that the spin current is generated in the GdIG layer, as there is no sign-change of a possible YIG-spin current expected. 
This sign change from the SP-generated spin current is compatible with the net magnetization orientation of the GdIG layer, which is inverted at the compensation temperature. In contrast, the spin current generated from SSE depends on the orientation of the Fe sublattices \cite{ohnuma_spin_2013,geprags_origin_2016,cramer_magnon_2017}. The orientation of the sublattices does not change across the compensation temperature due to the coupling to the YIG layer (see Fig. \ref{fig:Fig1} d)), which explains the absence of this sign-change in the SSE measurement of the bilayer compared to previous studies \cite{geprags_origin_2016,cramer_magnon_2017}. 

For GdIG single layers, at $T_{\mathrm{C,GdIG}}$ one can observe a divergence of the linewidth of the FMR signal. We can extract a linewidth from Fig. \ref{fig:Fig5} a), which clearly shows no such divergence of the linewidth in our case. This divergence was studied previously and linked to the relation between net angular momentum and total angular moment of the material \cite{ng_survey_2022, kamra_gilbert_2018}. However, the experimental investigation is difficult, due to the decrease in signal intensity and increasing linewidth. In our system, however, the relation between the net angular momentum and total angular momentum is shifted by the coupling to the YIG layer. Thus, close to the compensation temperature, we still observe a signal originating from the GdIG layer (Fig. \ref{fig:Fig5} a)). While the linewidth does change slightly across $T_{\mathrm{C,GdIG}}$, it is compatible with the damping estimated in Ref. \cite{ng_survey_2022}, supporting the approach of extracting the damping for ferrimagnetic materials close to their compensation temperature differently compared to ferromagnets.

\subsection{Spin Seebeck effect measurement}
The SSE measurements performed in the course of this study give important insight into the static magnetic state of the system for the respective applied external magnetic field as it follows the magnetization of our GdIG top layer \cite{becker_magnetic_2021,cramer_magnon_2017}.

We compare the SSE signal caused by microwave heating with the SSE signal from a simple continuous temperature gradient, to verify the field dependence of the SSE signal which is measured during the application of the FMR.
For the latter, the sample is clamped in between a cold sink, on which a Pt-stripe is attached, and a resistive Pt-heater \cite{cramer_magnon_2017}. When applying a current to the top Pt-heater, a temperature gradient is established perpendicular to the sample plane. The voltage which builds up on the sample is measured with an HP 34420A nanovoltmeter. 
For the SSE signal during the microwave application, we measure the $V_{\mathrm{iSHE}}$ during the field sweep. By applying a strong enough microwave power, we incidentally heat the film, additionally inducing the spin Seebeck effect \cite{an_unidirectional_2013,cheng_quantitative_2021}. This background signal is dependent on the magnetization direction of the GdIG layer, which gives us a direct comparison between the switching state of the top layer and the FMR signal. This signal includes the distinct peak of the SP at resonance (at $ \mu_0 \boldsymbol{\mathrm{H}} \approx$ \SI{0.15}{\tesla}), see Figure \ref{fig:Fig4} b). 
When we compare it to the DC-SSE, we can see a good agreement in the field dependence (apart from the aforementioned SP-signal peak at FMR). 

As observed in Ref. \cite{becker_magnetic_2021}, the SSE signal is dominated by the GdIG top layer. This means, that for the appropriate temperature range, we can see the switching during the field sweep of the GdIG layer. We do not observe a clear contribution of the YIG bottom layer, even if a superposition of spin currents could be apparent in any of our SSE measurements \cite{fan_manipulation_2020,cramer_magnon_2018}. It appears, that the spin current generated in the YIG layer cannot penetrate through the GdIG layer due to its larger damping and the interface between YIG and GdIG.

After comparison with the microwave-generated SSE, we can see good agreement with the DC-SSE signal. For further investigations, we mainly use the microwave-generated SSE. This enables a more meaningful comparison, as no remounting of the sample is needed.

A good indication, that the observed background signal during the microwave application indeed stems from the SSE originating from the GdIG, is the sign change at low temperatures. The signal vanishes at low temperature, which agrees with the sign change of the SSE in GdIG due to a change of occupied magnon modes (Fig. \ref{fig:Fig5} c))\cite{geprags_origin_2016}.  As the microwave heating power is dependent on the microwave frequency, for the applied constant microwave power, the frequency is kept constant at $f \approx$ \SI{6.5}{\giga\hertz} while sweeping the magnetic field.

\subsection{Simulation}
The data from our experiments are compared to micromagnetic simulations (Fig.\ref{fig:Fig4} a)). These were conducted using OOMMF via the Ubermag meta-package \cite{beg_ubermag_2022} written for python. The initial parameters are set by literature values. We assume the damping parameter $\alpha_{\mathrm{YIG}}$ to be \num{3e-3} which we extracted from single layer YIG measurements and compatible with PLD-grown YIG samples \cite{kumar_damping_2022}. The GdIG damping is estimated to be \num{5e-2} from \cite{ng_survey_2022}. The magnetization of each layer is estimated from literature values for a set temperature \cite{dionne_magnetic_2009}. 
An interfacial coupling between YIG and GdIG is assumed and estimated from the exchange constant, determined in \cite{becker_magnetic_2021}. The result of such a simulation is shown in Figure \ref{fig:Fig4} a). The overall signal shape is comparable to our FMR spectrum. However, in the simulation, we can see a higher order mode, which we could not resolve in our experimental data. The exchange-driven mode \cite{belmeguenai_frequency-_2007, drovosekov_magnetic_2020} with negative frequency to field dispersion seems also to be only a weak signal. However, the effect of its anti-crossing with the main mode is captured by the measurement (Fig. \ref{fig:Fig4} c)). 

\section{Conclusion}
In this study, we investigated the spin dynamics and spin currents in a  trilayer of YIG / GdIG / Pt.  With complementary measurements of the spin current from SSE and SP at FMR, and comparison to our micromagnetic simulations, we can explain the features of the FMR spectrum. 

We even observed the GdIG FMR of this system close to the compensation temperature $T_{\mathrm{C,GdIG}}$. The spin current, pumped at the ferromagnetic resonance, indicates the GdIG to be the origin of the spin current, which offers the opportunity to study the GdIG layer resonance close to its compensation temperature. The coupling to the YIG layer is found to prevent the resonance linewidth of the GdIG from diverging at $T_{\mathrm{C,GdIG}}$, in contrast to measurements of single-layer GdIG. 
We show that our system can be used as a highly tunable spin current source over a large temperature and external magnetic field range.


\begin{acknowledgments}
Funded by the Deutsche Forschungsgemeinschaft (DFG, German Research Foundation) - TRR 173/2 - 268565370 Spin+X (Projects B02, B13, and A01).
\end{acknowledgments}

\bibliography{Bib.bib}

\end{document}